\documentclass[11pt]{article}
\usepackage[T1]{fontenc}
\usepackage[utf8]{inputenc}
\usepackage{lmodern}
\usepackage[a4paper,margin=2.55cm]{geometry}
\usepackage{amsmath,amssymb,amsfonts}
\usepackage{graphicx}
\usepackage[numbers,sort&compress]{natbib}
\usepackage{authblk}
\usepackage{booktabs}
\usepackage{placeins}
\usepackage[font=small,labelfont=bf]{caption}
\usepackage{microtype}
\usepackage[colorlinks=true,allcolors=blue]{hyperref}

\newcommand{\rmd}{\mathrm{d}}
\title{Inverse reconstruction of Einstein cluster collapse and oscillations
in tangential-pressure spacetimes}
\author[1]{Jan J. Ostrowski\thanks{Jan.Jakub.Ostrowski@ncbj.gov.pl}}
\affil[1]{Department of Fundamental Research, National Centre for Nuclear
Research, Pasteura 7, 02--093 Warsaw, Poland}
\date{}

\begin{document}
\maketitle

\begin{abstract}
We formulate an inverse reconstruction procedure for spherically symmetric
Datta--Bondi--Gair spacetimes with vanishing radial pressure and 
non-zero tangential pressure supported by angular momentum. Extending the logic of Krasi\'nski--Hellaby reconstruction for
Lema\^itre--Tolman--Bondi dust, we determine the full, exact space-time information
from prescribed density, velocity or equilibrium state.
Two specific constructions are developed. The first represents an initially expanding fluid with a finite turnaround, followed by an
asymptotic collapse to an unstable Einstein cluster configuration. In the
second, a static equilibrium density and a finite time density-velocity pair
determine finite amplitude oscillations around a stable Einstein cluster. 
Numerical examples use a Jaffe-type final profile and an Einasto equilibrium with an approximate Navarro--Frenk--White finite time
profile. The latter admits many oscillations before the first shell crossing is
detected. The constructions provide diagnostics of the admissibility of prescribed configurations.  
\end{abstract}

\section{Introduction}
\label{sec:intro}

Spherical collapse models provide an analytically tractable setting for
studying gravitational instability, the development of inhomogeneity and the
limits of pressure support. Beginning with the Oppenheimer--Snyder dust
solution~\cite{OppenheimerSnyder1939}, methods based on spherically symmetric exact solutions have complemented perturbation theory and numerical investigations of cosmic
structure formation~\cite{Peebles1980LSSU,GunnGott1972,FillmoreGoldreich1984,
Bertschinger1985,2009BolejkoBook,2011Bolejko}. Although such models are not
substitutes for collisionless $N$-body or kinetic calculations, they make the
relation between prescribed matter distributions and spacetime geometry
particularly transparent. Examples of useful physical insights obtained through analytical (exact and approximate) solutions include: relation of the cosmic expansion to the local dynamics  (see e.g. \cite{CarreraGiulini}), upper limits on the largest gravitationally bound objects \cite{Pavlidou, 2022Ostrowski} and relativistic interpretation of the spiral galaxies' rotation curves \cite{Lake}.

Our main motivation for this paper is to develop an exact solution representing i) collapse scenarios in which the end-states are given by static configurations ii) models oscillating around static spacetime. The static configuration that we have in mind is described in General Relativity by spacetimes in which there exists a time-like Killing vector ensuring the conservation of energy. Stars, dark matter halos of galaxies and galaxy clusters can, to a reasonable degree, be approximated with the spherical symmetry. In such case, the static metric can be put in the following form
\begin{equation}
\label{eq:metric_static}
\begin{aligned}
\rmd s^2={}&-\mathrm{e}^{2\Phi(r)}\rmd t^2
+\left(1-\frac{2M(r)}{r}\right)^{-1}\rmd r^2+r^2\rmd\Omega^2,\\
\rmd\Omega^2={}&\rmd\theta^2+\sin^2\theta\,\rmd\phi^2.
\end{aligned}
\end{equation}
where $M(r)$ denotes the Misner--Sharp mass. The corresponding energy density
obeys
\begin{equation}
\label{eq:Misner-Sharp}
M(r)=4\pi\int_0^r\rho(\tilde r)\tilde r^2\,\rmd\tilde r.
\end{equation}
It is clear that such configurations for non-vacuum cases cannot solely rely on gravitational interaction  which is universally attractive - we need a physical component that will counter the gravity. If we restrict ourselves to the spherical symmetry and a diagonal stress-energy tensor, we can see that the pressure and the gradients of pressure can play a balancing role in static set-ups. Exact conditions for hydrodynamical equilibrium in spherical symmetry are given by the Tolman-Oppenheimer-Volkoff (TOV) equation \cite{Tolman1939, OppenheimerVolkoff1939}.
For matter with radial pressure $p_r$ and tangential pressure $p_T$, the
anisotropic TOV equation reads
\begin{equation}
\label{eq:TOV_general}
\frac{\rmd p_r}{\rmd r}=-(\rho+p_r)
\frac{M+4\pi r^3p_r}{r(r-2M)}+\frac{2(p_T-p_r)}{r}
\end{equation}
The possibility of support by
tangential stress, even when $p_r=0$, motivates dynamical extensions of the
Lema\^itre--Tolman--Bondi (LTB) dust spacetime. In a static case, such solutions are known as Einstein clusters.

An Einstein cluster (EC) is an idealized spherically symmetric ensemble of
counterrotating collisionless particles, whose angular motions generate
tangential pressure but no net bulk rotation~\cite{Einstein1939}.
Its vanishing radial velocity dispersion places it at the strongly
tangentially anisotropic limit of more general collisionless systems
\cite{BinneyTremaine,Merritt1985}. Einstein clusters have been revisited in various contexts,
from theoretical explorations of anisotropic fluids and relativistic stellar models, 
to phenomenological descriptions of dark-matter halos.
A recent work~\cite{Acharyya2023} has shown that realistic halo density laws such as the Einasto \cite{Einasto1965} or Navarro--Frenk--White (NFW, \cite{Navarro1997}) profiles can be accommodated within the Einstein cluster framework by suitable choices of the angular momentum distribution.

Dynamical, purely tangential-pressure models were examined by
Datta~\cite{Datta}, Bondi~\cite{BondiDatta}, and Gair~\cite{Gair2001}.
Related collapse and equilibrium questions were investigated by
Magli~\cite{Magli1997}
 and by Malafarina and Joshi~\cite{MalafarinaJoshi2010} in the context of naked singularities, while Joshi, Malafarina and Narayan~\cite{JoshiMalafarinaNarayan2011}
demonstrated that equilibrium configurations can arise as asymptotic states of tangential pressure collapse.
Related investigations of critical phenomena in EC collapse~\cite{Mahajan2007,Harada1998}
have highlighted the role of angular momentum in determining the final outcome of gravitational instability. Their results establish the possible shell dynamics, including circular
orbits and final states of evolution. In our complementary approach the question is instead
inverse: which shell and metric functions generate specified density,
velocity, equilibrium or asymptotic endpoint data? This extends the
reconstruction viewpoint developed by Krasi\'nski and Hellaby for LTB dust
\cite{HellabyKrasinski2002,KrasinskiHellaby,HellabyKrasinski2,KrasinskiHellaby2} and allows for full spacetime reconstruction and regularity diagnostics. 

We develop two constructions within the Datta--Bondi--Gair (DBG) family.
The first relates a prescribed unstable Einstein cluster endpoint to an
expanding outer branch that turns around and approaches the endpoint at
infinite particle proper time. The second prescribes a stable Einstein cluster
equilibrium and a finite-time density-velocity profile to construct
finite amplitude oscillations. Both require an additional coordinate
reconstruction: mass derivatives at fixed particle proper time are not in
general the physical Jacobian at fixed comoving time. We supply this
transformation and use it to test density positivity and occurence of shell crossing.

The article is organized as follows: 
section~\ref{sec:gair} recalls the DBG formalism, and
Sec.~\ref{sec:hk} the relevant LTB reconstruction method.
The two DBG inverse constructions and their admissibility conditions are
developed in Sec.~\ref{sec:HKDBG}. Section~\ref{sec:numerical} presents a
truncated Jaffe-type collapse example and an Einasto equilibrium with an
NFW-matched oscillating profile. We summarize the conclusions and possible
applications in Sec.~\ref{sec:conclusions}. Unless stated otherwise, we use
geometrized units $G=c=1$.

\section{Datta--Bondi--Gair solution}
\label{sec:gair}
The spherically symmetric models with anisotropic tangential pressure are well understood (\cite{Datta, BondiDatta, Evans, Magli1997, Magli2}). We will refer to these models as Datta--Bondi--Gair (DBG) models.  In this section we will closely follow the exposition from Gair (\cite{Gair2001}).  

We start by writing down a general, spherically symmetric metric in co-moving coordinates 
\begin{equation}
\label{eq:metric_spherical}
ds^2 = -\mathrm{e}^{2\nu (t,r)}\,\rmd t^2
       +  \mathrm{e}^{2\lambda (t,r)}\,\rmd r^2
       + R^2(t,r) \left(\rmd \theta^2 + \sin^2\theta\,\rmd \phi^2\right)\;,
\end{equation}
where $\nu$ and $\lambda$ are free functions and $R$ is the areal radius. The stress-energy momentum tensor has vanishing radial pressure and takes the following form  
\begin{equation}
T^{\mu}{}_{\nu}
= \mathrm{diag}\,[-\rho,\, 0,\, p_{T},\, p_{T}]\;,
\end{equation}
where $\rho$ is the energy density measured in the comoving frame and $p_T$ is the tangential pressure. The physical origin of this pressure is an effective centrifugal force generated by an ensemble of particles with angular momentum $L(r)$. At each point of the shell labelled with the coordinate $r$, the net vector angular momentum is by construction zero. Hence, the system remains spherically symmetric and non-rotating, while the scalar angular momentum is non-zero. Since this pressure is not of hydrodynamical origin, all the particles interact only gravitationally and follow geodesics. As a consequence, the angular momentum of each particle is a constant of motion and remains fixed
\begin{equation}
\label{eq:angular}
    L(r) = R^2(t,r)\;\sqrt{\left(\frac{\rmd  \theta}{\rmd \tau}\right)^2 +\;\sin^2 \theta \;\left(\frac{\rmd  \phi}{\rmd \tau}\right)^2} \;,
\end{equation}
where $\tau$ is the proper time of a constituent particle on the shell. 

Using the $(t,r)$ coordinates and denoting $\left(\partial/\partial t\right)_r$ with an overdot, the Hamiltonian constraint of this system reads
\begin{equation}
\label{eq:gair_ham}
\frac{1}{2} e^{-2\nu} \dot{R}^{2}
 =  \frac{M}{R} + \frac{\Lambda R^{2}}{6} - \frac{L^{2}}{2(L^{2}+R^{2})} + \frac{E R^{2}}{L^{2}+R^{2}}\;,
\end{equation}
where  $\Lambda$ is the cosmological constant, $M$ is the Misner-Sharp mass and $E=E(r)$ plays the role of a conserved energy per unit mass. To obtain a closed form solution, Gair chooses to use the $(\tau, r)$ coordinates, where $\tau$ is the proper time of a constituent particle, and thus Eq. ~\eqref{eq:gair_ham} takes the following form:
\begin{equation}
\label{eq:gair_ham2}
\frac{1}{2}\left(\frac{\partial R}{\partial \tau}\right)^2_r
 = E + \frac{M}{R}\left(1+\frac{L^2}{R^2}\right) +\frac{\Lambda}{6}\left(L^2+R^2\right) -\frac{L^2}{2R^2}\;.
\end{equation}
Before we proceed further, it is worth mentioning another natural choice of the time parameter, $T$, which is the time experienced by an observer at rest on the fixed shell (no angular momentum). In that case, the Hamiltonian constraint reads
\begin{equation}
\label{eq:gair_ham3}
\frac{1}{2}\left(\frac{\partial R}{\partial T}\right)_r^{2}
 =  \frac{M}{R} + \frac{\Lambda R^{2}}{6} - \frac{L^{2}}{2(L^{2}+R^{2})} + \frac{E R^{2}}{L^{2}+R^{2}}\;.
\end{equation}
This might sometimes be a more convenient choice in the context of the reconstruction of the spacetime which will be described in the following sections. For the remainder of this article we will assume zero cosmological constant $\Lambda=0$.

Equations \eqref{eq:gair_ham}, \eqref{eq:gair_ham2} and \eqref{eq:gair_ham3} can be thought of as one-dimensional equations of motion for a particle in the potential with the centrifugal barrier. To further utilize this analogy, we rewrite Eq.~\eqref{eq:gair_ham2} using the potential $V = -\frac{M}{R}\left(1+\frac{L^2}{R^2}\right) +\frac{L^2}{2R^2}\; \;,$ which provides a more compact form for our equation,
\begin{equation}
\label{eq:hamiltonian_potential}
  \frac{1}{2}\left(\frac{\partial R}{\partial \tau}\right)^2_r = E - V.  
\end{equation}
The radial equation can be
integrated shell by shell
\begin{equation}
\label{eq:quadrature}
A(R,r)=\int_0^R \frac{\rmd\tilde R}
{\sqrt{2[E(r)-V(\tilde R,r)]}}=\tau-\tau_B(r).
\end{equation}
The integration function $\tau_B$ is a bang-time function.
For an oscillating shell, the origin of the quadrature must instead be a
turnaround point within the allowed radial interval. For the bound case an elliptic integral representation is given in Appendix~\ref{Appa}. The explicit form of the areal radius $R(\tau, r)$ can be obtained by the inversion of $A(R,r) = \tau-\tau_B(r)$. Eq. ~\eqref{eq:quadrature} together with partial differential relations between two sets of coordinates, $(t,r)$ and $(\tau, r)$ allow Gair to write down the metric in $(\tau,r)$ coordinates, with all of the functions in coefficients expressed in a closed form. The coordinate-transformation relations needed to reconstruct the full metric are made explicit in Sec.~\ref{sec:oscillation}. What is important in the context of this work is that the Datta--Bondi--Gair model is described by 4 free functions, $(M(r), E(r), L(r), \tau_B(r))$ of which three independent functions remain after fixing the radial shell-label freedom. 

There are six possible shell evolution types, depending on the value of $E(r)$ and $V(r)$: i) expansion and re-collapse, ii) bouncing universe, iii) expansion to infinity or collapse from infinity, iv) oscillating solution, v) Einstein cluster, vi) coasting universe. This is considerably more than in the LTB spacetime, and each case requires a slightly different approach. 

\subsection{Einstein cluster}
\label{sec:einstein_cluster}
Solution ~\eqref{eq:metric_spherical} admits a static limit which is the Einstein cluster corresponding to the $\dot{R}=0$ and $\ddot{R}=0$ conditions. This translates to
\begin{equation}
\label{eq:ec_conditions}
E(r) = V(R,r) \;\;,\;\;
\frac{\partial V}{\partial R} = 0.
\end{equation}
We can use the radial gauge freedom to relabel shells $R(r) \to r$ and infer from Eq.~\eqref{eq:ec_conditions} the following formulas for $L$ and $E$,
\begin{equation}
\label{eq:ec_relations}
L = \sqrt{\frac{rM}{1 - \frac{3M}{r}}} \;\;,\;\;E = \frac{1}{2}\left(-1 + \frac{\left(1 - \frac{2M}{r}\right)^2}{1 - \frac{3M}{r}}\right)\;.
\end{equation}
Circular configurations on the branch $R_0>6M$ are stable against
small radial perturbations, whereas $3M<R_0<6M$ is the unstable branch.
A nontrivial orbit may approach a circular radius asymptotically only when
the radial potential has an accessible double root with $V''(R_{\rm EC})<0$.
Near this root, $E-V\propto(R-R_{\rm EC})^2$, and the proper-time
quadrature diverges logarithmically. Consequently, the asymptotic endpoint
of Sec.~\ref{sec:ec_final_state} must lie on the unstable branch; an orbit
near the stable minimum instead oscillates. A stationary shell remains
at equilibrium, but a generic nearby shell does not converge to it in
the absence of dissipation.

It was shown in \cite{JoshiMalafarinaNarayan2011} that the single model of Einstein cluster is an asymptotic state (infinite proper time) of an infinite number of initial density configurations. This is true because the information contained in density profiles alone can not fix the space-time uniquely. The authors propose a closure condition by specifying pressure component - this however is not very practical as knowledge of pressure requires an implicit knowledge of the areal radius via
\begin{equation}
\label{pressure}
    p(r,t) = \frac{1}{2}\rho(r,t)\frac{L(r)^2}{L(r)^2+R^2(t,r)} \;.
\end{equation}
Our reconstruction scheme allow us to fix the geometry from data on two hypersurfaces or one finite-time hypersurface and end-state. 

It is worth noting that the vanishing of the radial pressure makes the
DBG family particularly well suited to modelling bounded objects. Let
$\Sigma$ be the timelike comoving hypersurface $M=M_{\rm tot}$ bounding
the configuration, with outward unit normal $n^{\mu}$. The Darmois
conditions for matching \eqref{eq:metric_spherical} to an exterior across
$\Sigma$ without a surface layer require the normal-normal and
normal-tangential projections of the stress-energy tensor to vanish
there,
\begin{equation}
T_{\mu\nu}n^{\mu}n^{\nu}\big|_{\Sigma}=p_{r}=0\ ,\qquad
T_{\mu\nu}n^{\mu}e^{\nu}_{(a)}\big|_{\Sigma}=0\ ,
\label{eq:junction}
\end{equation}
the second holding identically because the frame is comoving and there
is no radial energy flux. Since $p_{r}\equiv0$ throughout the DBG
spacetime, \eqref{eq:junction} is satisfied on every comoving
shell. A smooth match to a Schwarzschild exterior of mass
$M_{\rm tot}$ is possible when the induced metric and extrinsic
curvature are matched across $\Sigma$. No condition is
imposed on $\rho$ or on $p_{T}$ at $\Sigma$, so the density and the
tangential pressure may jump there without generating a surface layer.
This is in contrast to perfect fluid collapse, where the matching
requires the pressure itself to vanish on the boundary, and a finite outer boundary can be smoothly matched to an
asymptotically flat Schwarzschild exterior. This matching does not
remove the singular centre of the collapse example. Matching instead to a
cosmological background, which additionally constrains the boundary to
comove with that background, is left for future work.

\subsection{Oscillating solution}
We can perturb the Einstein cluster to obtain an oscillating solution around the static configuration. Following \cite{Gair2001} we keep $M$ and $L$ fixed and perturb $R \to r_{EC}+\delta R (r,t)$, $\nu \to \nu_{EC}+\delta \nu(r,t) $ and $E \to E_{EC}+\delta E(r)$. The Hamiltonian constraint then gives us
\begin{equation}
    \frac{\partial^2 \delta R}{\partial \tau^2} = -\frac{M(R_0-6M)}{R_0^3(R_0-3M)} \delta R\;,
\end{equation}
which is a harmonic oscillator equation with a frequency given by
\begin{eqnarray}
    \omega = \sqrt{\frac{M\left(R_0-6M\right)}{R^3_0(R_0-3M)}} \;.
\end{eqnarray}
General solution reads
\begin{equation}
    R(\tau,r) = R_0 + A(M) \cos \left(\omega(M)(\tau-\tau_{\phi}(M))\right) \;,
\end{equation}
where $A$ is the amplitude and $\tau_{\phi}$ is the phase. Since this spacetime is dynamical, there is a possibility of occurrence of shell crossing i.e. $R'(t,r) = 0$. To avoid them, one has to impose a condition that all the shells have the same period of oscillations in the $t$ parameter. However, as will be shown, without this condition certain choices of density profiles allow for several oscillation periods before the shell crossing occurs. It is another reason why the full metric reconstruction is important for diagnostics of the given solution.

\section{The Krasi\'nski--Hellaby reconstruction method}
\label{sec:hk}
In this section, we will briefly summarize the main features of the LTB metric and the Krasi\'nski--Hellaby method of reconstructing this spacetime from the data on two distinct hypersurfaces ~\cite{HellabyKrasinski2002}.

The LTB metric in synchronous co-moving coordinates reads
\begin{equation}
    \label{eq:ltb-metric}
\rmd s^2 = -\rmd t^2 +\frac{\left(R'\right)^2}{1+2E}\rmd r^2 +  R^2 \left(\rmd \theta^2 + \sin^2\theta\,\rmd \phi^2\right)\;, 
\end{equation}
where $R=R(t,r)$ is again the areal radius and $E(r)$ is the total energy of the shell per unit mass.

The field equations read (${}':= \partial/\partial r$)
\begin{equation}
\label{eq:field-ltb}
    4\pi \rho = \frac{M'}{R^2R'} \;\;,\;\;\dot{R}^2 = \frac{2M}{R}+2E \;\;.
\end{equation}
Equation for the areal radius can be obtained in a parametric form depending on the sign of the $E$ function
\begin{eqnarray}
\label{eq:ltb_parametric}
&E<0& \;\;,\;\;R=-\frac{M}{2E}\left(1-\cos \mu\right) \;\;,\;\;\mu - \sin \mu =\frac{\left(-2E\right)^{3/2}}{M}\left(t-t_B\right) \\
&E=0&\;\;,\;\;R=\left(\frac{9}{2}M\left(t-t_B\right)^2\right)^{1/3} \\
&E>0& \;\;,\;\;R=\frac{M}{2E}\left(\cosh \mu -1\right)\;\;,\;\;\sinh \mu -\mu = \frac{\left(2E\right)^{3/2}}{M}\left(t-t_B\right) \;,
\end{eqnarray}
where $\mu$ is an auxiliary parameter and $t_B$ is another integration function called the bang time.

The LTB spacetime is thus described by three free functions $(M(r), E(r), t_B(r))$  two of which are sufficient to fully determine the spacetime. From a practical standpoint however, it is more convenient to use density or velocity profiles which at given instance of time can be inferred from observations. In a series of papers ~\cite{HellabyKrasinski2002, KrasinskiHellaby, HellabyKrasinski2, KrasinskiHellaby2} Krasi\'nski and Hellaby developed a reconstruction scheme that allows to recover all the LTB's free functions from certain combinations of data on two distinct hypersurfaces. These include density-to-density, velocity-to-velocity and (density-)velocity-(velocity-)density reconstructions and more.
Below we will briefly sketch the density-to-density evolution for the recollapsing ($E<0$) case. 
We start by noting that it is convenient to use $M(r)$ function itself
as a radial coordinate (this assumes a good behaviour for $M(r)$, i.e., growing, no singularities).
With that coordinate choice we have
\begin{equation}
\label{eq:M_coordinate}
    4\pi \rho = \frac{3}{(R^3)_{,M}} \;\;,\;\;R^3-R^3(M_0) = \int_{M_0}^{M}\frac{3}{4\pi \rho(\tilde{M})} \;\rmd \tilde{M}.
\end{equation}
Let us pick two hypersurfaces at $t_1$ and $t_2$, with two density profiles $\rho_1$ and $\rho_2$ respectively. Using Eq. ~\eqref{eq:M_coordinate} at two time instances and the solution for $E<0$ from Eq. ~\eqref{eq:ltb_parametric}  we get
\begin{equation}
\label{eq:HK_radius}
R_i(M) = R(t_i,M)=-\frac{M}{2E}\left(1-\cos \mu\right) \;\;,\;\;\mu - \sin \mu =\frac{\left(-2E\right)^{3/2}}{M}\left(t_i-t_B\right)\;\;,\;\;i=1,2 \;.    
\end{equation}
 From here on, the reconstruction procedure depends on whether our second instance of time, $t_2$, is still in the expanding phase or already in the collapsing one. Since we are mostly concerned with the final stages of collapse we will present the latter option which requires inverse of cosine in the range $\mu \in [\pi, 2\pi]$. For $t=t_1$ we write down the bang function and then insert it into Eq. ~\eqref{eq:HK_radius} for $t_2$ which results in
\begin{eqnarray}
\label{eq:KH_method}
    0 &=& \pi + \arccos\left(-1+a_2x\right) +\sqrt{1-\left(1-a_2x\right)^2} \nonumber\\
    &-&\arccos\left(1-a_1x\right)+\sqrt{1-\left(1-a_1x\right)^2} - \left(t_2-t_1\right)\;x^{3/2} \;,
\end{eqnarray}
where
\begin{equation*}
    a_i = \frac{R_i}{M^{1/3}} \;\;,\;\;x = -\frac{2E}{M^{2/3}} \;\;.
\end{equation*}
Solving this equation for $E(M)$ and using the previous relation for $t_B(M)$ we determine two out of the three free functions and hence determine the geometry of the entire space time. The KH procedure does not guarantee the physical viability of a solution. This has to be checked separately after the unique solution to Eq. ~\eqref{eq:KH_method} is found. For instance, the solution should not contain any shell crossings, marked by $R'=0$ condition, nor should it contain too negative total energy $E>-1/2$. We will expand the list of viability conditions once we generalize the KH method in the next section.

\section{The KH method for DBG solutions}
\label{sec:HKDBG}
In this and the following section we present the main body of our work. We aim at the extension of the KH method to anisotropic systems (in terms of pressure) described by the DBG metrics. Similar to the case in the original method, the main motivation is to reconstruct the spacetime from astrophysically-driven data. None of the four functions describing the DBG spacetime, $(M,E,L, \tau_B)$, is easily or directly accessible via the observations. These functions are also less physically intuitive than the density or velocity profiles.   

Before we proceed, let us summarize schematically the KH method for the density-to-density evolution.
\begin{itemize}
    \item We start by choosing two hypersurfaces, at $t_1$ and $t_2$, and two density profiles, $\rho_1$ and $\rho_2$.
    \item By using $M$ as a radial coordinate and utilizing Eq.~\eqref{eq:M_coordinate}, we find $R_1(M)$ and $R_2(M)$.
    \item The energy function $E(M)$ is then found by integrating the equation of motion
from $R_1(M)$ to $R_2(M)$ when the radial velocity has a fixed sign:
\begin{equation}
\label{eq:hk_energy2}
\sigma(t_2-t_1)
 = \int_{R_{1}}^{R_{2}}
   \frac{\rmd \tilde{R}}
        {\sqrt{2\left(E(M)-V_{\rm LTB}(M, \tilde{R})\right)}}.
\end{equation}
where $\sigma=+1$ for expansion and $\sigma=-1$ for collapse,
and $V_{\rm LTB}=-M/R$ is understood in this dust equation.
If the shells turn around between the two hypersurfaces, the time
of flight must instead be split at $R_{\rm ta}=-M/E$:
\[
t_2-t_1=\int_{R_1}^{R_{\rm ta}}\frac{\rmd R}{\sqrt{2(E-V_{\rm LTB})}}
+\int_{R_2}^{R_{\rm ta}}\frac{\rmd R}{\sqrt{2(E-V_{\rm LTB})}}.
\]
\item Once $E(M)$ is known, the bang-time function $t_B(M)$ follows from the expanding branch by
\begin{equation}
\label{eq:hk_tb}
t_{B}(M) = t_{1} - \int_{0}^{R_{1}}
 \frac{\rmd \tilde{R}}{\sqrt{2\left(E(M)-V_{\rm LTB}(M, \tilde{R})\right)}}.
\end{equation}
\end{itemize}
Equations~\eqref{eq:hk_energy2} and~\eqref{eq:hk_tb}
provide a closed reconstruction scheme:
data on two space-like hypersurfaces determine the geometry of spacetime entirely.

When attempting to extend the KH method to the DBG solutions, one quickly realizes two main differences: the number of free functions needed to make the problem determined is three instead of two, i.e., we have to fix three functions from $(M, E, L,  \tau_B)$. Moreover, Eq. ~\eqref{eq:hk_energy2} now contains an extra term because of the angular momentum. These make the extension of KH method non-trivial. The first problem is rather easily circumvented by requiring one more function on any of the hypersurfaces (e.g. velocity or angular momentum profile). The second problem is coming from the fact that the solution to Eq.~\eqref{eq:quadrature} is an elliptic integral, whose explicit form relies on the roots of the cubic equation, which is known only after the whole reconstruction is performed (see ~\ref{Appa}). This difficulty however can be overcome through numerical procedures.

\subsection{Reconstruction of the collapsing Einstein cluster}
\label{sec:ec_final_state}
In this section, we will present the step-by-step algorithm for the DBG spacetimes reconstruction with the final stage described by the Einstein cluster. This model represents initially expanding structure that undergoes the turnaround and approaches EC asymptotically.

For each fixed shell $M$, the radial equation can be written as
\begin{equation}
\label{eq:W}
\dot R^2 = W(R,M)=2E+\frac{2M}{R}-\frac{L^2}{R^2}+\frac{2ML^2}{R^3}.
\end{equation}
Here and in the remainder of this section a dot denotes differentiation with respect to the shell proper time $\tau$.
It is useful to remove the denominators by introducing the cubic polynomial
\begin{equation}
    P(R) \equiv R^3 W(R,M)= 2ER^3+2MR^2-L^2R+2ML^2 ,
\end{equation}
and its derivative
\begin{eqnarray}
    P'(R) = 6ER^2+4MR-L^2,
\end{eqnarray}
A shell can approach a static Einstein-cluster radius $R_{\rm EC}(M)$ asymptotically only if $R_{\rm EC}$ is a repeated root of $P$. Thus
\begin{equation}
P(R_{\rm EC})=0,
\qquad
P'(R_{\rm EC})=0.
\end{equation}
Solving these two equations gives
\begin{equation}
L^2(M)=\frac{M R_{\rm EC}^2}{R_{\rm EC}-3M},
\end{equation}
and
\begin{equation}
E(M)=
\frac{M(4M-R_{\rm EC})}
{2R_{\rm EC}(R_{\rm EC}-3M)}.
\end{equation}
 The repeated root condition is the algebraic expression of the fact that the proper time integral diverges logarithmically near the Einstein cluster radius:
a simple root would be reached in finite shell time, whereas a double root is approached only asymptotically.

The third root is obtained by factorizing
\begin{equation}
P(R)=2E(R-R_{\rm EC})^2(R-R_3).
\end{equation}
For the branch relevant to collapse with a finite turnaround we require
\begin{equation}
4M<R_{\rm EC}<6M,
\end{equation}
which is consistent with requiring  for the final orbits to be unstable ($3M<R_{\rm EC}<6M$). 
To see this, let us recall that at the circular radius,
\begin{equation}
W(R_{\rm EC},M)=0,
\qquad
W_{,R}(R_{\rm EC},M)=0.
\end{equation}

Around the circular orbit we can expand
\begin{equation}
    W(R_{EC}+\epsilon) \approx \frac{1}{2}W''(R_{EC})\epsilon^2 + \mathcal{O}(\epsilon^3)
\end{equation}
For a nontrivial solution approaching the double root from either side,
the leading quadratic term must make $W$ positive in the allowed region.
Thus $W_{,RR}(R_{\rm EC})>0$. Since $W=2(E-V)$, this is equivalent to
$V_{,RR}(R_{\rm EC})<0$. The asymptotic EC radius therefore corresponds
to a maximum of the effective potential, i.e. to the unstable circular branch.

In the $(4M,6M)$ interval $E<0$, and the third root, representing the turnaround radius, lies outside $R_{\rm EC}$. We denote it by
\begin{equation}
R_{\rm ta}(M)=R_3=\frac{2M R_{\rm EC}}{R_{\rm EC}-4M}.
\end{equation}
The radial equation then takes the factorized form
\begin{equation}
W(R,M)=\frac{-2E(M)}{R^3}\left(R-R_{\rm EC}(M)\right)^2\left(R_{\rm ta}(M)-R\right).
\end{equation}
Since $E<0$, this expression is non-negative for $0<R<R_{\rm ta}$, except at the double root $R=R_{\rm EC}$.

 The double root separates the motion into two disconnected sectors. In the
inner sector, $0<R<R_{\rm EC}$, a shell may start from $R=0$ at finite shell
time and approach $R_{\rm EC}$ asymptotically from below. This branch has no
turnaround. In the outer sector,
\begin{equation}
R_{\rm EC}<R\le R_{\rm ta},
\end{equation}
a shell may start arbitrarily close to $R_{\rm EC}$ in the infinite past,
expand to $R_{\rm ta}$, and then collapse back toward $R_{\rm EC}$ asymptotically.
In what follows we use this outer branch.

Near the double root,
\begin{equation}
\label{eq:double_root}
W(R,M)\simeq \kappa^2(M)\left(R-R_{\rm EC}\right)^2,
\end{equation}
where
\begin{equation}
\kappa^2(M)
=
\frac{M(6M-R_{\rm EC})}
{R_{\rm EC}^3(R_{\rm EC}-3M)}.
\end{equation}
Thus
\begin{equation}
R-R_{\rm EC}\sim e^{-\kappa |\tau|},
\end{equation}
which makes explicit the asymptotic character of the final state.

The reconstruction therefore proceeds as follows. The target EC density fixes
$R_{\rm EC}(M)$ and hence $E(M)$, $L(M)$, and $R_{\rm ta}(M)$ through the
relations above. The initial density profile at $\tau=\tau_1$ fixes
$R_1(M)$. For the outer-branch construction we impose
\begin{equation}
R_{\rm EC}(M)<R_1(M)<R_{\rm ta}(M).
\end{equation}
Assuming that the shell is initially expanding, the proper-time distance from
$R_1$ to the turnaround is
\begin{equation}
\Phi(R_1,M)
=
\int_{R_1}^{R_{\rm ta}}
\frac{dR}{\sqrt{W(R,M)}}.
\end{equation}
The turnaround time of each shell is therefore
\begin{equation}
\tau_{\rm ta}(M)=\tau_1+\Phi(R_1(M),M).
\end{equation}
The full outer-branch evolution is then obtained by solving
\begin{equation}
\label{eq:inversion}
\Phi(R,M)=|\tau-\tau_{\rm ta}(M)|
\end{equation}
for $R(\tau,M)$, where
\begin{equation}
\Phi(R,M)
=
\int_{R}^{R_{\rm ta}}
\frac{d\bar R}{\sqrt{W(\bar R,M)}}.
\end{equation}
The expanding and collapsing branches are distinguished by the sign of
$\dot R$:
\begin{equation}
\dot R>0 \quad \text{for} \quad \tau<\tau_{\rm ta}(M),
\end{equation}
and
\begin{equation}
\dot R<0 \quad \text{for} \quad \tau>\tau_{\rm ta}(M).
\end{equation}
It is important that the final EC radius is not reached at a finite value of
$\tau$. The hypersurface $R=R_{\rm EC}$ is therefore not a second finite-time
Cauchy surface in the ordinary KH sense. The role normally played by the
bang time is here played by $\tau_{\rm ta}(M)$, the
shell-dependent turnaround time. This is why the reconstruction differs from
the standard density-to-density KH problem: one input is an asymptotic
configuration rather than a finite-time density profile.

A numerical realization of this construction is presented in Sec.~\ref{subsec:numerics_GHK}.
\subsection{Reconstruction of the oscillating solution}
\label{sec:oscillation}
We now consider the stable branch of the Einstein cluster configuration and
construct finite-amplitude oscillations around it. The input consists of

\begin{enumerate}
    \item a static density profile $\rho_0$, which determines the equilibrium
    radius $R_0(M)$,
    \item a density profile $\rho_1$ on a finite-time hypersurface
    $\tau=\tau_1$, which determines $R_1(M)$,
    \item a radial velocity profile $U_1(M)=\dot R(\tau_1,M)$ on the same
    hypersurface.
\end{enumerate}
Equivalently, instead of prescribing $U_1(M)$ one could prescribe a second
finite-time density profile $R_2(M)$ at a known time separation
$\tau_2-\tau_1$. In that case $E(M)$ would be found shell by shell from the
time-of-flight integral between $R_1$ and $R_2$. In the present construction
we use the velocity profile because the equilibrium profile $R_0(M)$ is not a
second finite-time hypersurface; it fixes the potential minimum around which
the shell oscillates.
The first profile fixes the stable EC background. In mass gauge, $\rho_0$
is inverted to obtain $R_0(M)$, and the circular orbit conditions give
\begin{equation}
L_0^2(M)=\frac{M R_0^2}{R_0-3M},
\end{equation}
and
\begin{equation}
E_0(M)=
\frac{M(4M-R_0)}
{2R_0(R_0-3M)} .
\end{equation}
The stable branch requires
\begin{equation}
R_0(M)>6M .
\end{equation}

The second density profile gives $R_1(M)$, while the velocity profile gives
the energy of the actual oscillating shell
\begin{equation}\label{eq:osc_energy}
E(M)
=
\frac{U_1^2}{2}
-\frac{M}{R_1}
-\frac{M L_0^2}{R_1^3}
+\frac{L_0^2}{2R_1^2}.
\end{equation}
Thus the static profile fixes the potential well, whereas the pair
$(R_1,U_1)$ fixes the particular orbit in that well. A density profile alone
would not determine the spacetime, since it would not specify whether the shell
is moving inward or outward, nor its oscillation amplitude.

For each shell, the turning points are obtained from
\begin{equation}
W(R,M)
=
2E(M)+\frac{2M}{R}
-\frac{L_0^2(M)}{R^2}
+\frac{2M L_0^2(M)}{R^3}
=0 .
\end{equation}
Equivalently, one solves the cubic
\begin{equation}
P(R)
=
R^3 W(R,M)
=
2ER^3+2MR^2-L_0^2R+2ML_0^2 .
\end{equation}
For a bound oscillation around the stable Einstein cluster equilibrium, the three positive roots can be ordered as follows
\begin{equation}
R_i<R_{\rm peri}<R_{\rm apo},
\end{equation}
where $\rm peri$ and $\rm apo$ stand for pericentre and apocentre respectively. 
The stable oscillation around the Einstein cluster equilibrium occurs in the outer allowed interval
\begin{equation}
R_{\rm peri}\le R\le R_{\rm apo}.
\end{equation}
The smallest root $R_i$ bounds a separate inner allowed region. It is
separated from the stable oscillation region by the forbidden interval
$R_i<R<R_{\rm peri}$, where $W<0$ and hence $\dot R^2<0$ would be required.

For $E<0$ the radial equation can be factorized as
\begin{equation}
W(R,M)
=
\frac{-2E(M)}{R^3}
\left(R-R_i\right)
\left(R-R_{\rm peri}\right)
\left(R_{\rm apo}-R\right),
\end{equation}
which makes the sign structure explicit. The stable oscillating branch is
therefore the outer well between $R_{\rm peri}$ and $R_{\rm apo}$.

The proper time from pericentre to radius $R$ on the outward branch is given by integral
\begin{equation}
I(R,M)
=
\int_{R_{\rm peri}}^{R}
\frac{d\bar R}{\sqrt{W(\bar R,M)}} .
\end{equation}
The half-period and full period are
\begin{equation}
H(M)=I(R_{\rm apo},M),
\qquad
T(M)=2H(M).
\end{equation}

The initial phase is determined by the sign of $U_1$. Let
\begin{equation}
I_1(M)=I(R_1(M),M).
\end{equation}
If $U_1>0$, the shell is on the outward branch and
\begin{equation}
s_1(M)=I_1(M).
\end{equation}
If $U_1<0$, the shell is on the inward branch and
\begin{equation}
s_1(M)=T(M)-I_1(M).
\end{equation}
At a turning point $U_1=0$, the phase is fixed by whether $R_1=R_{\rm peri}$ or $R_1=R_{\rm apo}$. The pericentre time is therefore
\begin{equation}
\tau_{\rm peri}(M)=\tau_1-s_1(M).
\end{equation}

At an arbitrary time $\tau$, define the phase time within one radial period,
\begin{equation}
s(\tau,M)
=
\left[\tau-\tau_{\rm peri}(M)\right]\ {\rm mod}\ T(M),
\qquad
0\le s<T.
\end{equation}
For the outward half-cycle, $0\le s\le H$, the areal radius is obtained by
solving
\begin{equation}
I(R,M)=s .
\end{equation}
For the inward half-cycle, $H<s<T$, the radius is obtained by solving
\begin{equation}
I(R,M)=T-s .
\end{equation}
The sign of the velocity is then
\begin{equation}
\dot R=
\begin{cases}
+\sqrt{W(R,M)}, & 0<s<H,\\[4pt]
-\sqrt{W(R,M)}, & H<s<T .
\end{cases}
\end{equation}
Once $R(\tau,M)$, $E(M)$ and $L_0(M)$ are known, the metric and stress-energy
components are recovered from
\begin{equation}
\label{eq:metricfns}
    {\rm e}^{2\lambda} = \frac{R^2_{,M}}{1+2E}\left(1+\frac{L^2}{R^2}\right) \;,\;\nu_{,M} = \frac{L^2}{R^2+L^2}\frac{R_{,M}}{R}\;,\;\left(\frac{\partial \tau}{\partial t}\right)_{r} = \frac{{\rm e}^{\nu}}{\sqrt{1+\frac{L^2}{R^2}}}\;,
\end{equation}
and Eq. ~\eqref{pressure}. 

\paragraph{Transformation to proper-time coordinates.}

An important distinction must be made between radial derivatives at fixed
comoving time $t$ and at fixed particle proper time $\tau$.
The derivatives entering Eq.~\eqref{eq:metricfns} are evaluated at fixed $t$,
whereas the shell trajectories obtained from the proper time quadratures
are functions of the form of $R(\tau,M)$.

Following Gair~\cite{Gair2001}, we introduce
\begin{equation}
K(\tau,M)
\equiv
\left(\frac{\partial\tau}{\partial M}\right)_t,
\qquad
U(\tau,M)
\equiv
\left(\frac{\partial R}{\partial\tau}\right)_M.
\end{equation}
The chain rule gives
\begin{equation}
\label{eq:radial_derivative_transform}
D(\tau,M)
\equiv
\left(\frac{\partial R}{\partial M}\right)_t
=
\left(\frac{\partial R}{\partial M}\right)_\tau
+
U(\tau,M)K(\tau,M).
\end{equation}

The function $K$ is determined by the compatibility of the
coordinate transformation with Eq.~\eqref{eq:metricfns}.
Differentiating $(\partial\tau/\partial t)_M$ with respect to
$M$ at fixed $t$, and using the relation for $\nu_{,M}$,
we obtain
\begin{equation}
\label{eq:K_evolution}
\left(\frac{\partial K}{\partial\tau}\right)_M
=
\frac{2L^2}{R(R^2+L^2)}D
-
\frac{L L_{,M}}{R^2+L^2}.
\end{equation}
This is an evolution form of the coordinate-transformation
relation derived by Gair.

We synchronize the origins of particle proper time on the
initial comoving hypersurface $t=t_1$, choosing
\begin{equation}
\label{eq:K_initial}
K(\tau_1,M)=0.
\end{equation}
Equations~\eqref{eq:radial_derivative_transform}
and~\eqref{eq:K_evolution} then determine $K$ and $D$
once the shell trajectories $R(\tau,M)$ are known.

Writing
\begin{equation}
q^2(\tau,M)=1+\frac{L^2(M)}{R^2(\tau,M)},
\end{equation}
the complete metric in $(\tau,M)$ coordinates becomes
\begin{equation}
\label{eq:metric_proper_time}
ds^2
=
-q^2\left(d\tau-K\,dM\right)^2
+
\frac{q^2 D^2}{1+2E}\,dM^2
+
R^2d\Omega^2.
\end{equation}
Unlike the original comoving metric, this expression
generally contains a mixed $d\tau\,dM$ term.

Finally, in the mass coordinate the energy density is
\begin{equation}
\label{eq:rho_corrected}
\rho(\tau,M)
=
\frac{1}{4\pi R^2(\tau,M)D(\tau,M)}.
\end{equation}
For $M>0$ and $R>0$, absence of shell crossings requires
\begin{equation}
\label{eq:shell_crossing_corrected}
D(\tau,M)>0.
\end{equation}
Thus positivity of $(\partial R/\partial M)_\tau$
alone is not sufficient to establish regular shell ordering.

\subsection{Consistency checks for reconstructed DBG metrics}

The reconstruction procedure fixes the free functions of the DBG solution, but
it does not by itself guarantee that the resulting spacetime is physically
admissible over the desired time interval. The following checks are imposed
after the reconstruction.

First, the metric coefficient $e^{2\lambda}$ must remain real and non-singular.
In the mass gauge this requires
\begin{equation}
1+2E(M)>0,
\end{equation}
together with regularity of the factor $1+L^2/R^2$ for the range of radii
considered. Hence
\begin{equation}
E(M)>-\frac12 .
\end{equation}

Second, the mass shells must remain well ordered. As shown in
Eqs.~\eqref{eq:radial_derivative_transform} and
\eqref{eq:rho_corrected}, the physical energy density is
determined by the radial Jacobian
$D=(\partial R/\partial M)_t$.

Positivity of the energy density and absence of shell crossings
therefore require
\begin{equation}
D(\tau,M)>0
\end{equation}
throughout the considered spacetime domain.
 No separate energy conditions need to be imposed. With
$T^{\mu}{}_{\nu}=\mathrm{diag}[-\rho,0,p_{T},p_{T}]$ and $p_{T}$ given by
\eqref{pressure}, write
\begin{equation}
p_{T}=w\,\rho\ ,\qquad
w(\tau,M)=\frac{L^{2}}{2\bigl(L^{2}+R^{2}\bigr)}\in\Bigl(0,\tfrac12\Bigr)\ .
\label{eq:w}
\end{equation}
Then, wherever $\rho\ge0$,
\begin{alignat}{2}
&\text{NEC:}&\quad \rho+p_{r}&=\rho\ge0\ ,\qquad
 \rho+p_{T}=(1+w)\rho\ge0\ , \nonumber\\
&\text{WEC:}&\quad \rho&\ge0 \ \text{ together with the NEC}\ ,
 \nonumber\\
&\text{DEC:}&\quad \rho&\ge|p_{r}|=0\ ,\qquad
 \rho\ge|p_{T}|=w\rho\ \ \text{since } w<\tfrac12\ , \nonumber\\
&\text{SEC:}&\quad \rho+p_{r}+2p_{T}&=(1+2w)\rho\ge0\ .
\label{eq:ec}
\end{alignat}
All four conditions therefore hold automatically, and the strict
inequality $w<1/2$ means the tangential pressure never exceeds
$\rho/2$. Consequently, on a domain with $R>0$, the condition
$D>0$, which already guarantees $\rho>0$ and the absence of shell
crossings, is the only matter condition that has to be monitored in the
reconstruction.

Third, the reconstructed shell motion must be consistent with the intended
branch of the effective potential. For the asymptotic collapse-to-EC
construction, the endpoint must lie on the unstable circular branch,
\begin{equation}
4M<R_{\rm EC}(M)<6M,
\end{equation}
and the initial data must be placed on the outer branch,
\begin{equation}
R_{\rm EC}(M)<R_1(M)<R_{\rm ta}(M).
\end{equation}
These conditions ensure that the shell has a finite turnaround radius and then
approaches $R_{\rm EC}$ asymptotically.

For the oscillating construction, the equilibrium radius must lie on the stable
circular branch,
\begin{equation}
R_0(M)>6M.
\end{equation}
The energy reconstructed from $(R_1,U_1)$ must place the shell in the outer
stable well
\begin{equation}
E_0(M)<E(M)<{\rm min}(E_u(M),0)~,
\end{equation}
where $E_u(M)$ is the potential value at the unstable circular orbit with the
same $M$ and $L_0(M)$. Equivalently, the turning points must satisfy
\begin{equation}
R_i<R_{\rm peri}<R_1<R_{\rm apo}
\end{equation}
for an initially outward moving shell, with the corresponding reversed branch
allowed when $U_1<0$.

Finally, the time evolution must be checked over the intended interval. In the
oscillating case, even if the initial phases are synchronized, different shell
periods can lead to dephasing and eventually to shell crossing. Therefore
$D>0$ is not assumed from the initial data; it is verified numerically on
the evolved slices used in the examples below.

Only after these checks have been performed can the reconstructed functions be
interpreted as defining an admissible DBG spacetime on the selected time
domain.

\section{Numerical Examples}
\label{sec:numerical}

To illustrate the reconstruction scheme we present two numerical examples. The
first realizes the outer branch construction of Sec.~\ref{sec:ec_final_state}, in which a set
of shells expands to a finite turnaround radius and then collapses
asymptotically toward an unstable Einstein cluster configuration. The second
example realizes the oscillating construction of Sec.~\ref{sec:oscillation},
in which an Einasto profile sets the stable EC equilibrium and an
NFW-matched profile is imposed at finite proper time.

The examples are not intended as realistic simulations of halo formation. With $\Lambda=0$ the construction contains no dimensionful constant
beyond those carried by the solution itself. The radial equation
\eqref{eq:W}, the metric functions \eqref{eq:metricfns} and the matter
relations \eqref{pressure} are consequently form-invariant
under the one-parameter homothety
\begin{equation}
M\to kM\ ,\quad R\to kR\ ,\quad L\to kL\ ,\quad \tau\to k\tau\ ,\quad
 \rho\to k^{-2}\rho\ ,\quad p_{T}\to k^{-2}p_{T}\ ,
\label{eq:scaling}
\end{equation}
for any $k>0$, with $E\to E$, under which $\mathrm{d}s^{2}\to k^{2}\,\mathrm{d}s^{2}$. The overall mass scale of an example is therefore a
choice of units and carries no physical information; the physical
content is expressed by dimensionless compactness, angular momentum,
density-shape and phase profiles. The overall dimensional scale is arbitrary. The examples are
accordingly chosen so that relativistic effects are visible at modest
numerical cost, not to represent observed halos; by \eqref{eq:scaling}
the algorithm itself is unchanged at realistic compactness, only the
numerical dynamic range increases. Their
purpose is to demonstrate that the reconstruction algorithm produces explicit
DBG spacetimes from density and velocity data, and to show how the resulting
solutions can be checked for admissibility.
\subsection{Asymptotic collapse to an Einstein cluster}
\label{subsec:numerics_GHK}

For practical applications the Krasi\'nski--Hellaby reconstruction
is most conveniently implemented shell by shell in the comoving
mass coordinate $M$.
In this subsection we outline a minimal numerical scheme that connects
the initial density profile to late Einstein cluster
regime.

For the asymptotic endpoint, we choose a truncated Jaffe-type mass law \cite{Jaffe}. Specifically
\begin{equation}
    M(R) = \frac{M_J R}{a_J+R} \;,
\end{equation}
where $M_J$ and $a_J$ are constant. The EC radius reads
\begin{equation}
    R_{\rm EC} = \frac{a_JM}{M_J-M} = Q(M)M\;,
\end{equation}
where auxiliary function $Q(M)$ has to belong to $(4,6)$ interval for the solution to be consistent (unstable orbit).
The endpoint density associated with this mass law is
\begin{equation}
    \rho_{\rm EC} = \frac{\left(M_J-M\right)^4}{4\pi a_J^3M_J M^2} \;. 
\end{equation}
As $M\to0$, $R_{\rm EC}\propto M$ and
$\rho_{\rm EC}\propto M^{-2}$, giving a genuine singular centre.
The numerical reconstruction therefore excludes $M=0$ and describes
only the outer matter region; no globally regular centre is claimed.

Other relevant quantities in terms of function $Q(M)$ read
\begin{equation}
    E(M) = \frac{4-Q}{2Q(Q-3)} \;,\;L^2(M)=\frac{Q^2M^2}{Q-3}\;,\; R_{\rm ta}(M) = \frac{2QM}{Q-4} \;.
\end{equation}
The initial profile is obtained from the following construction
\begin{equation}
\label{eq:profile}
    R_1 = S(M)R_{\rm EC} \;,\;S(M) = 1.18+0.05\exp\left(-\left(\frac{M-0.45}{0.2}\right)^2\right)\;.
\end{equation}
Since $S(M)>1$ and
\begin{equation}
    S(M)<\frac{R_{\rm ta}}{R_{\rm EC}} = \frac{2}{Q(M)-4},
\end{equation}

we have $R_{\rm EC}<R_1<R_{\rm ta}$. The density profile takes the form
\begin{equation}
    \rho_1(M) = \frac{\rho_{\rm EC}(M)}{S^2(M)\left(S(M)+S'(M)\frac{M(M_J-M)}{M_J}\right)} \;.
\end{equation}
The integral describing the time from the initial slice to turnaround is
\begin{equation}
\label{phase}
\Phi(R_1,M)
=
\int_{R_1}^{R_{\rm ta}}
\frac{dR}{\sqrt{W(R,M)}} .
\end{equation}
The endpoint $R=R_{\rm ta}$ is a turnaround point. For numerical evaluation
we use the substitution
\begin{equation}
\bar R(u)=R_{\rm ta}-Au^2,
\qquad
A=R_{\rm ta}-R_1,
\qquad
0\le u\le1 .
\end{equation}
This gives
\begin{equation}
\label{eq:ec_phase}
\Phi(R_1,M)
=
2\sqrt{\frac{A}{-2E}}
\int_0^1
\frac{\bar R(u)^{3/2}}
{\bar R(u)-R_{\rm EC}}
\,du .
\end{equation}
For the inversion of Eq.~\eqref{eq:inversion}, the same expression is used with
$R_1$ replaced by the unknown radius $R$.

The numerical procedure is then as follows. First, $R_{\rm EC}(M)$, $E(M)$,
$L^2(M)$ and $R_{\rm ta}(M)$ are computed on a uniform mass grid, excluding the
central point $M=0$ where the chosen profile has a cusp. Second, the initial
radius $R_1(M)$ is constructed from Eq.~\eqref{eq:profile}, and the corresponding
turnaround time is obtained from
\[
\tau_{\rm ta}(M)=\tau_1+\Phi(R_1(M),M).
\]
The proper time evolution is characterized by the inversion relation
$\Phi(R,M)=|\tau-\tau_{\rm ta}(M)|$; numerically we integrate
its equivalent radial evolution equations as described in Sec.~\ref{sec:numerical_results}.
The physical shell Jacobian $D$ is obtained from
Eqs.~\eqref{eq:radial_derivative_transform} and~\eqref{eq:K_evolution}.
The energy density and single-stream condition are then given by
Eqs.~\eqref{eq:rho_corrected} and~\eqref{eq:shell_crossing_corrected},
respectively.

In the numerical example we set
\begin{equation}
M_{\rm tot}=1,\qquad M_J=6,\qquad a_J=27.
\end{equation}
Hence
\begin{equation}
Q(M)=\frac{27}{6-M},
\end{equation}
so that, on $0<M<1$,
\begin{equation}
4.5<Q(M)<5.4.
\end{equation}
The endpoint therefore lies entirely on the unstable EC branch
$4M<R_{\rm EC}<6M$.
\begin{figure}[!tbp]
    \centering
    \includegraphics[width=0.93\linewidth]
    {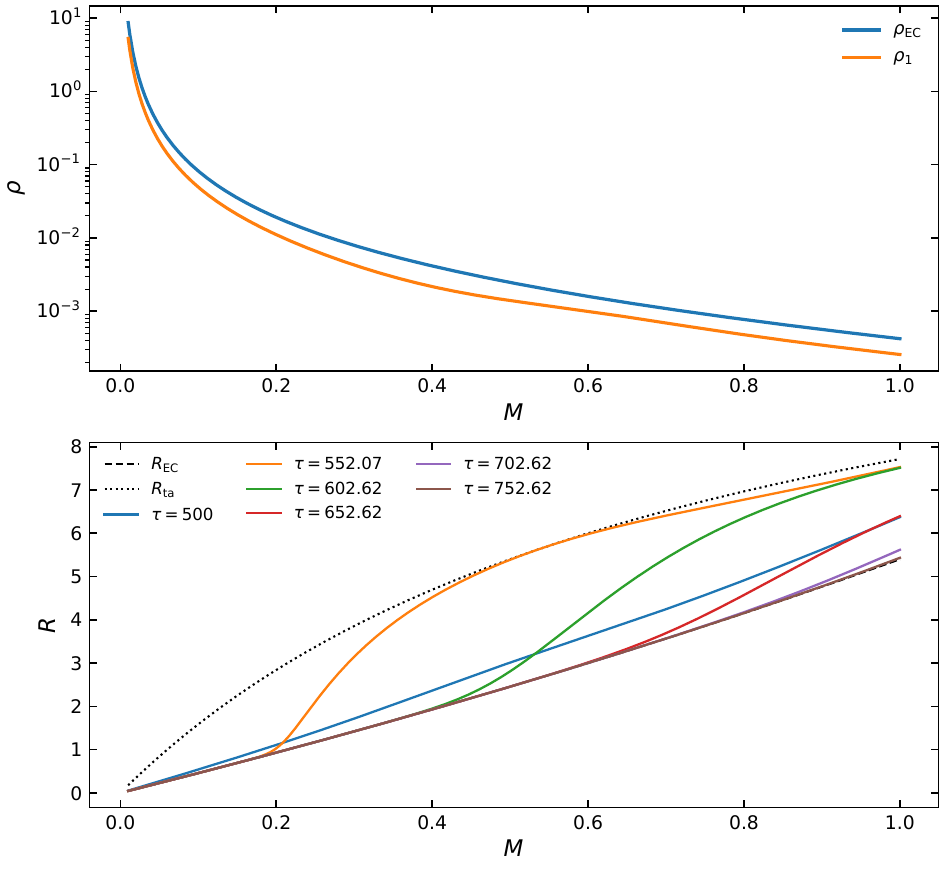}

    \caption{
    Reconstruction of the asymptotic collapse-to-Einstein-cluster
    solution. Upper panel: the prescribed Einstein-cluster density
    profile $\rho_{\rm EC}(M)$ and the initial density
    $\rho_1(M)$ obtained from the stretched configuration.
    Lower panel: evolution of the areal radius $R(\tau,M)$
    at selected proper times, together with the asymptotic
    Einstein-cluster radius $R_{\rm EC}(M)$ and the
    shell-dependent turnaround radius $R_{\rm ta}(M)$.
    Throughout the numerical examples, $M_{\rm tot}=1$.
    }

    \label{fig:collapse_input_evolution}
\end{figure}
\begin{figure}[!tbp]
    \centering
    \includegraphics[width=0.93\linewidth]
    {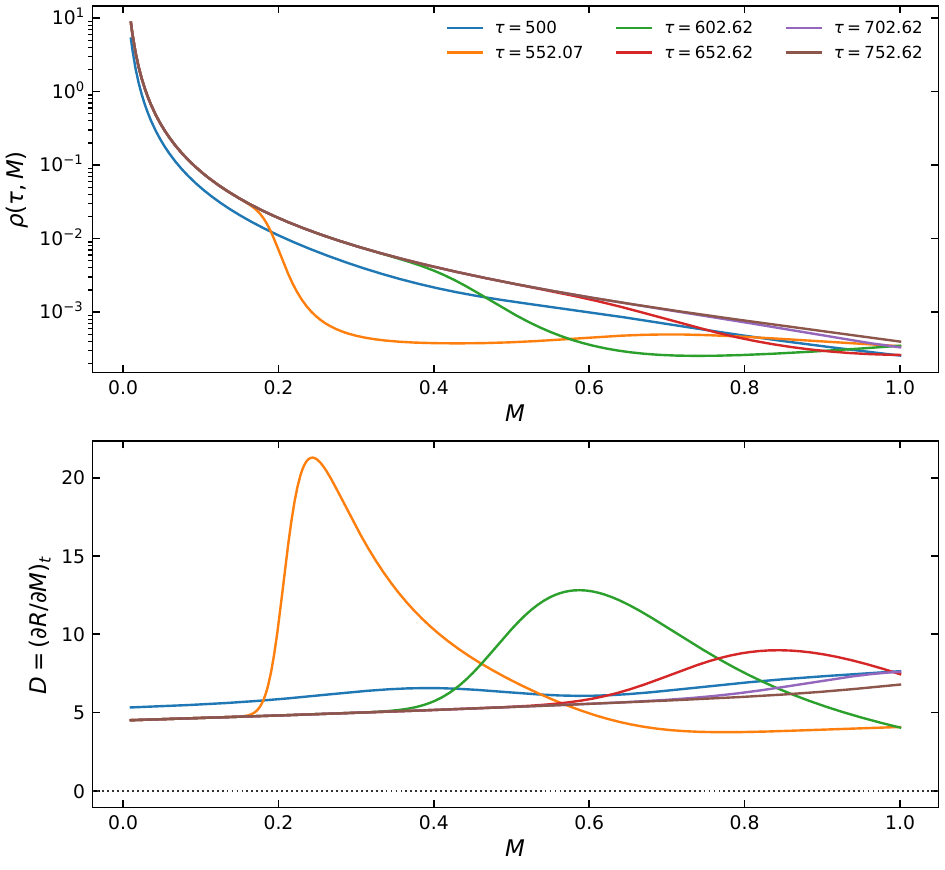}

    \caption{
    Density evolution and shell-separation diagnostic for the
    asymptotic collapse-to-Einstein-cluster reconstruction.
    Upper panel: energy density $\rho(\tau,M)$ evaluated using
    the physical radial Jacobian
    $D=(\partial R/\partial M)_t$.
    Lower panel: the corresponding Jacobian $D(\tau,M)$.
    The curves are evaluated at selected particle proper times.
    The condition $D>0$ is satisfied on all displayed slices
    over the sampled mass domain.
    }

    \label{fig:collapse_density_D}
\end{figure}
\subsection{Oscillating Einasto--NFW reconstruction}

For the oscillating example we choose the equilibrium density profile to be
Einasto,
\begin{equation}
\rho_{\rm Ein}(R)
=
\rho_{-2}
\exp\left[
-\frac{2}{\alpha}
\left(
\left(\frac{R}{R_{-2}}\right)^\alpha-1
\right)
\right],
\end{equation}
and use an NFW law to construct the outer part of the finite-time density profile,
\begin{equation}
\rho_{\rm NFW}(R)
=
\frac{\rho_s}{(R/R_s)(1+R/R_s)^2}.
\end{equation}
Both profiles are truncated at $R_{\rm vir}$ and normalized to the same total
mass $M_{\rm tot}$.
The equilibrium configuration is fixed by the Einasto density law, with
$\alpha=0.18$. We denote by $R_{\rm NFW}(M)$ the inverse cumulative mass
profile of the truncated, normalized NFW density law. To reduce the rapidly
varying central displacement while preserving the NFW profile in the outer
region, we prescribe the initial radius as
\begin{equation}\label{eq:matched_initial}
 R_1(M)=R_0(M)+w(M)\left[R_{\rm NFW}(M)-R_0(M)\right].
\end{equation}
The transition function is defined by
\begin{equation}\label{eq:matched_weight}
 w(M)=\epsilon+(1-\epsilon)S\!\left(\frac{M-M_a}{M_b-M_a}\right),
 \qquad
 S(x)=\begin{cases}
 0,&x\le0,\\
 6x^5-15x^4+10x^3,&0<x<1,\\
 1,&x\ge1.
 \end{cases}
\end{equation}
We take $\epsilon=0.01$, $M_a=0.02M_{\rm tot}$, and
$M_b=0.18M_{\rm tot}$. The resulting profile is $C^2$ in the
transition region. In particular, $R_1=R_{\rm NFW}$, and hence
$\rho_1=\rho_{\rm NFW}$, for $M\ge M_b$; the inner shells are
close to the Einasto equilibrium. This is an NFW-matched finite-time
slice rather than an exactly NFW profile over the full mass range.
The initial velocity remains
$U_1=-\eta\omega(M)[R_1(M)-R_0(M)]$, with $\eta=0.5$.

In the numerical example we use
\begin{equation}
M_{\rm tot}=1,\qquad R_{\rm vir}=100,\qquad c=8,
\end{equation}
with
\begin{equation}
R_s=R_{-2}=\frac{R_{\rm vir}}{c}=12.5,
\qquad
\alpha=0.18.
\end{equation}
The cumulative mass profiles are numerically inverted to obtain
$R_0(M)=R_{\rm Ein}(M)$ and $R_{\rm NFW}(M)$.
The finite-time radius is prescribed separately by
Eq.~\eqref{eq:matched_initial}.

\begin{figure}[!tbp]
    \centering
    \includegraphics[width=0.93\linewidth]
    {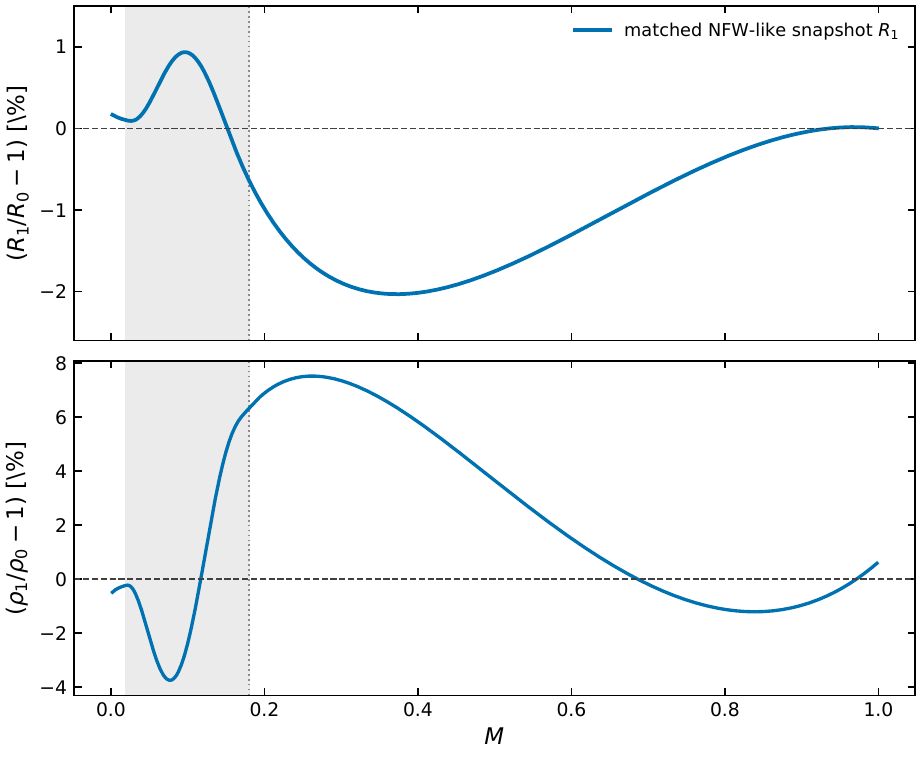}

    \caption{
    Comparison of the input configurations used in the
    oscillating reconstruction, evaluated at equal enclosed
    mass $M$.
    Upper panel: relative difference between the initial
    areal radius $R_1(M)$ and the Einasto equilibrium
    radius $R_0(M)$, expressed as a percentage.
    Lower panel: the corresponding relative difference
    between the initial density $\rho_1(M)$ and the
    equilibrium density $\rho_0(M)$.
    The shaded region, $0.02<M<0.18$, denotes the
    smooth transition between the near-equilibrium inner
    configuration and the NFW-matched outer profile.
    For $M\geq0.18$, the initial density coincides
    with the NFW profile.
    }

    \label{fig:oscillating_input_contrasts}
\end{figure}

The radial velocity on the NFW-matched slice is chosen as
\begin{equation}
U_1(M)=\dot R(\tau_1,M)
=
-\eta\,\omega(M)\left[R_1(M)-R_0(M)\right].
\end{equation}
In the harmonic approximation,
\begin{equation}
R-R_0=A\cos\Theta,\qquad
\dot R=-A\omega\sin\Theta,
\end{equation}
so that
\begin{equation}
U_1=-\omega(R_1-R_0)\tan\Theta.
\end{equation}
A constant $\eta$ therefore fixes the harmonic phase tangent $\tan\Theta=\eta$, with phases defined modulo $\pi$; it does not synchronize the exact nonlinear periods.

After $E(M)$ is reconstructed from Eq.~\eqref{eq:osc_energy}, we check that each shell remains
in the stable outer well,
\begin{equation}
E_0(M)<E(M)<E_u(M),
\end{equation}
where $E_u(M)$ is the value of the effective potential at the unstable circular
orbit for the same $M$ and $L_0(M)$. We also verify that the turning points
satisfy
\begin{equation}
R_i<R_{\rm peri}<R_1<R_{\rm apo}.
\end{equation}
\paragraph{Numerical implementation.}

Both examples are reconstructed on a uniform grid of 500 mass
shells. For the collapsing solution we use $0.01\leq M\leq 1$,
while for the oscillating solution the computational domain is
$0.002\leq M\leq 0.998$. The cumulative Einasto and NFW mass
profiles are evaluated by numerical quadrature and inverted
using monotone cubic interpolation.

The shell evolution is computed by integrating the radial
equations together with their mass derivatives, using an
adaptive Runge--Kutta method (RK45). For the collapse example,
a parameterization adapted to the double-root potential is
employed to preserve the asymptotic branch. The
coordinate-transformation function $K(\tau,M)$ is integrated
simultaneously, with the initial condition
$K(\tau_1,M)=0$. The physical radial Jacobian is then
evaluated as
$D=R_{,M}|_\tau+\dot R K$, and the energy density is
calculated from $\rho=(4\pi R^2D)^{-1}$.

The relative integration tolerances are $2\times10^{-9}$
and $3\times10^{-9}$ for the collapsing and oscillating
solutions, respectively, with an absolute tolerance of
$2\times10^{-10}$ in both cases. Shell crossings are
detected by locating the first zero of $\min_M D(\tau,M)$
during the numerical integration.

For the oscillating example, the first detected crossing
occurs at $\tau\simeq10462.46$ on the 500-shell grid.
Increasing the resolution to 1000 shells gives
$\tau\simeq10457.20$, a relative difference of approximately
$0.05\%$. These results indicate numerical convergence
of the detected crossing time over the sampled mass domain,
but do not constitute a proof of regularity in the continuum
limit.
\FloatBarrier
\subsection{Results and diagnostics}
\label{sec:numerical_results}

\paragraph{Asymptotic collapse to an Einstein cluster.}

The first reconstruction illustrates the evolution toward a prescribed
unstable Einstein cluster configuration. Figure~\ref{fig:collapse_input_evolution}
shows the density profiles of the asymptotic EC endpoint and the initial
configuration, together with the reconstructed areal-radius evolution.

Both density profiles exhibit a central cusp, inherited from the
Jaffe-type mass distribution. In particular, since
$R_{\rm EC}(M)\propto M$ near the centre, the endpoint density behaves
as $\rho_{\rm EC}\propto M^{-2}$. The numerical construction therefore
excludes the singular central point $M=0$.

The initial configuration is obtained by stretching the EC radius
according to $R_1(M)=S(M)R_{\rm EC}(M)$. For the chosen stretching
function, the initial density is lower than the endpoint density
over the considered mass range.

The lower panel of Fig.~\ref{fig:collapse_input_evolution} illustrates
the subsequent shell evolution. The curves $R_{\rm EC}(M)$ and
$R_{\rm ta}(M)$ define the asymptotic endpoint and the
shell-dependent turnaround radius, respectively.

The reconstructed shells follow the outer branch of the
double-root solution. Each shell initially expands, reaches
its turnaround radius, and subsequently collapses toward
$R_{\rm EC}(M)$. The endpoint is approached asymptotically
in the particle proper time and is never reached at a finite
value of $\tau$.

The corresponding energy-density evolution and regularity
diagnostics are presented in Fig.~\ref{fig:collapse_density_D}.
Unlike the radial derivative at fixed particle proper time, the physical
Jacobian $D=(\partial R/\partial M)_t$ determines both the energy
density and the shell crossing condition [Eqs.~\eqref{eq:rho_corrected}
and~\eqref{eq:shell_crossing_corrected}].

The upper panel of Fig.~\ref{fig:collapse_density_D} shows the
resulting density evolution. The moving features are associated
with the shell-dependent turnaround times and the resulting
redistribution of matter.

The lower panel shows the physical shell-separation diagnostic
$D(\tau,M)$. The condition $D>0$ is satisfied on all six
displayed slices. The smallest value over the displayed slices is approximately
$3.75>0$ (Table~\ref{tab:numerical_diagnostics}).
Thus no shell crossing is detected on the sampled slices.
This numerical test does not, by itself, establish regularity
at every intermediate time.

\paragraph{Oscillations around a stable Einstein cluster.}

The second reconstruction describes finite-amplitude oscillations
around a stable EC equilibrium. The equilibrium density follows
the Einasto profile, while the prescribed finite-time configuration
is smoothly matched to an NFW density distribution outside the
inner region.

Figure~\ref{fig:oscillating_input_contrasts} compares the two
input configurations through the relative differences in their
areal radii and energy densities at equal enclosed mass.

These differences are difficult to distinguish in plots of the
absolute profiles, since the corresponding curves nearly overlap.
The relative representation makes the deviations from equilibrium
more apparent.

The initial radius is constructed by smoothly connecting an
approximately equilibrated inner region to the NFW configuration.
For $M\geq0.18$, the initial density coincides with the NFW
profile. For $M\leq0.02$, the configuration remains close to
the Einasto equilibrium, while $0.02<M<0.18$ defines the
transition region.

The resulting initial profile is therefore NFW-matched rather
than globally NFW. This modification reduces the amplitude
of the inner shell perturbations and substantially extends
the interval over which the reconstructed configuration
remains free of shell crossings.

The reconstructed energies satisfy the bound orbit conditions
$E_0(M)<E(M)<E_u(M)$ to numerical precision. The minimum
separation from the unstable barrier is about $1.17$, the nearly
circular shells require tolerance-aware evaluation (Table~\ref{tab:numerical_diagnostics}).
The smallest difference $E-E_0$ is of order $10^{-14}$,
corresponding to an almost circular shell. The strict
inequality at this numerical precision should be interpreted
with the corresponding tolerance.

\begin{figure}[!tbp]
    \centering
    \includegraphics[width=0.93\linewidth]
    {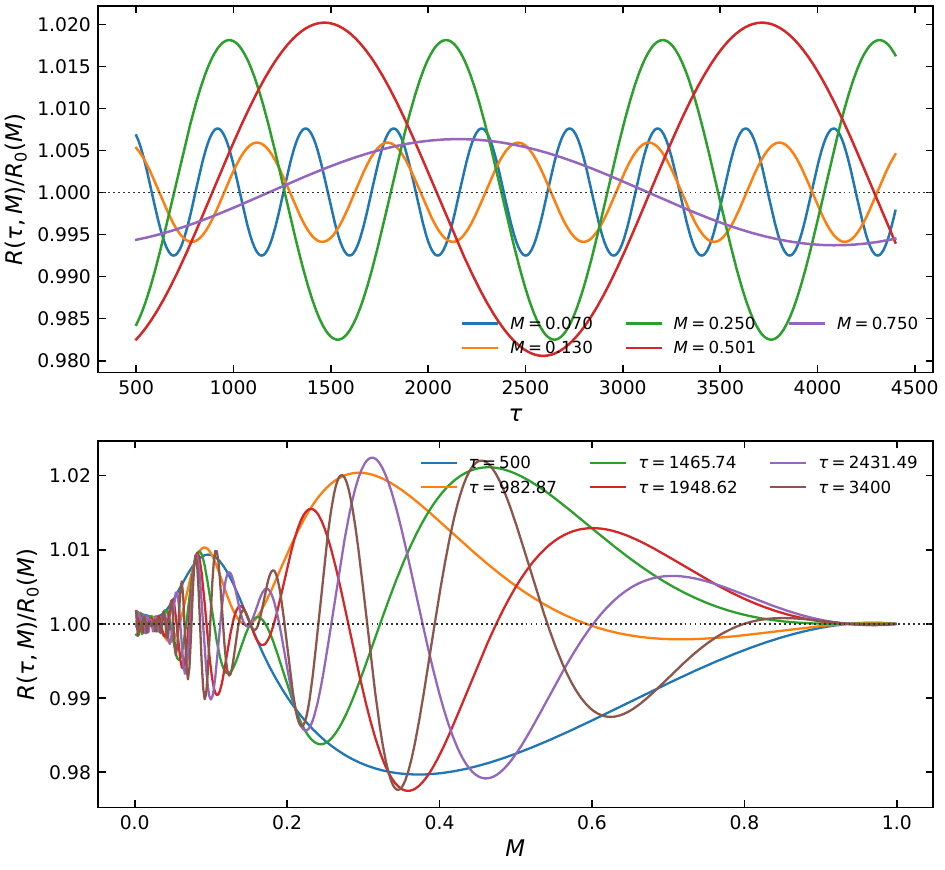}

    \caption{
    Radial evolution of the reconstructed oscillating
    Einstein-cluster configuration.
    Upper panel: trajectories of selected mass shells,
    showing oscillations around their equilibrium
    radii $R_0(M)$ over an extended proper-time interval.
    Lower panel: the fractional areal radius
    $R(\tau,M)/R_0(M)$ at selected proper times.
    The mass dependence of the oscillation period produces
    progressive dephasing between neighbouring shells.
    The displayed evolution remains within the numerically
    verified single-stream interval.
    }

    \label{fig:oscillating_kinematics}
\end{figure}

Figure~\ref{fig:oscillating_kinematics} illustrates the
resulting dynamics.

The upper panel shows the proper time evolution of selected
mass shells. The shells undergo sustained radial oscillations
around their respective equilibrium radii $R_0(M)$.
The trajectories are obtained from the exact nonlinear
radial equation rather than its harmonic approximation.

Consequently, the turning points are generally not symmetric
around $R_0(M)$, and the exact oscillation periods differ
from their linearized values. The lower panel of Fig.~\ref{fig:oscillating_kinematics}
shows the fractional radius $R(\tau,M)/R_0(M)$ at selected
proper times.
The radial displacement varies across the mass distribution
because both the oscillation amplitude and the period depend
on $M$. The progressive difference between the oscillation
phases of neighbouring shells produces the characteristic
spatial structure visible in the later profiles.
The numerical trajectories shown in the upper panel extend
to $\tau=4400$, while the discrete radial profiles in the
lower panel extend to $\tau=3400$.

\begin{figure}[!tbp]
    \centering
    \includegraphics[width=0.93\linewidth]
    {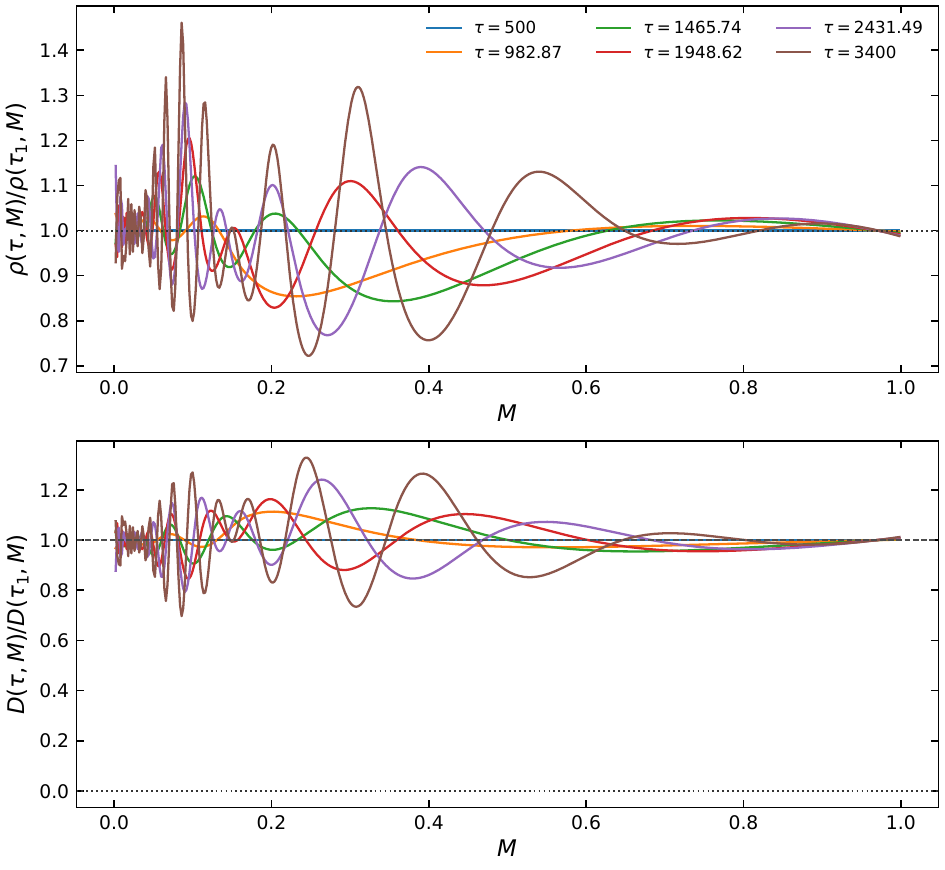}

    \caption{
    Evolution of the energy density and shell-separation
    diagnostic in the oscillating reconstruction.
    Upper panel: density relative to its initial value,
    $\rho(\tau,M)/\rho(\tau_1,M)$.
    Lower panel: the physical radial Jacobian
    $D=(\partial R/\partial M)_t$, normalized by its
    initial value $D(\tau_1,M)$.
    Both quantities are evaluated at events corresponding
    to the indicated particle proper times.
    Since $D(\tau_1,M)>0$, positivity of the normalized
    Jacobian is equivalent to $D(\tau,M)>0$.
    No shell crossings are detected on the displayed
    slices within the sampled mass domain.
    }

    \label{fig:oscillating_diagnostics}
\end{figure}

The corresponding evolution of the energy density and the
physical shell-separation diagnostic is shown in
Fig.~\ref{fig:oscillating_diagnostics}.
To make the relatively small variations more apparent,
we present both quantities relative to their values on the
initial hypersurface:
\begin{equation}
\frac{\rho(\tau,M)}{\rho(\tau_1,M)},
\qquad
\frac{D(\tau,M)}{D(\tau_1,M)}.
\end{equation}

The upper panel of Fig.~\ref{fig:oscillating_diagnostics}
shows the density perturbations generated by the oscillating
motion. Their spatial structure becomes progressively more
complex as neighbouring shells accumulate different phases.
The lower panel shows the normalized physical Jacobian.
Since $D(\tau_1,M)>0$, the sign of this ratio determines
whether shell ordering remains regular.
At the last displayed snapshot, $\tau=3400$, the minimum
Jacobian is $50.35>0$ (Table~\ref{tab:numerical_diagnostics}).
The Jacobian also remains positive on all earlier displayed
slices. Hence no shell crossing occurs at the sampled events
shown in Fig.~\ref{fig:oscillating_diagnostics}.

\FloatBarrier
\paragraph{Long-term regularity of the oscillating solution.}

\begin{figure}[!tbp]
    \centering
    \includegraphics[width=0.93\linewidth]
    {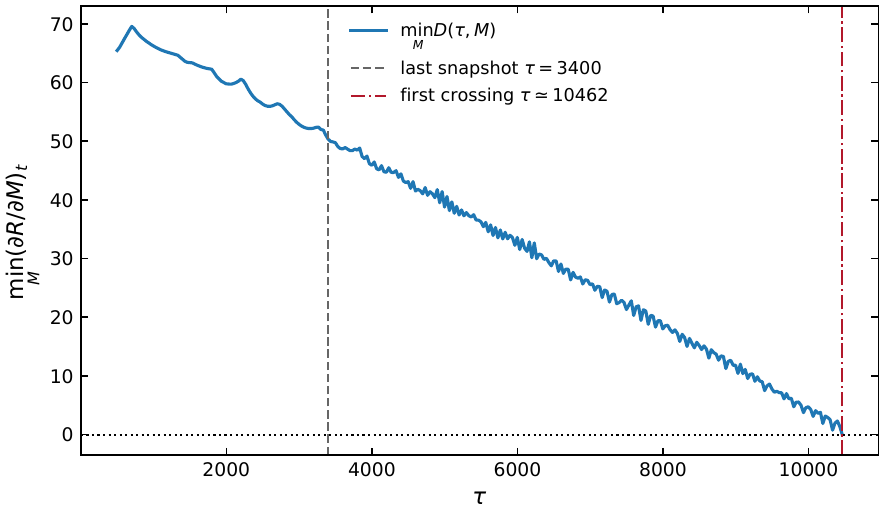}

    \caption{
    Long-term shell-crossing diagnostic for the oscillating
    Einstein-cluster reconstruction.
    The curve shows the minimum physical radial Jacobian
    $D=(\partial R/\partial M)_t$ over the sampled mass domain
    as a function of particle proper time.
    The vertical marker at $\tau=3400$ indicates the latest
    snapshot displayed in the density and radial-profile
    figures.
    The first shell crossing detected on the numerical grid
    occurs at approximately $\tau=10462$, when the minimum
    Jacobian reaches zero.
    The quoted time is resolution-dependent and should not
    be interpreted as an exact continuum crossing time.
    }

    \label{fig:oscillating_crossing_scan}
\end{figure}

The positivity of the radial Jacobian on selected time slices
does not guarantee regular shell ordering throughout the
entire evolution. In particular, the different oscillation
periods of neighbouring shells can eventually lead to
shell crossings.
We therefore extend the numerical evolution beyond the
interval displayed in the preceding figures and monitor
the minimum physical Jacobian over the sampled mass domain.
The result is shown in
Fig.~\ref{fig:oscillating_crossing_scan}.
For the adopted initial conditions, the physical Jacobian
remains positive over a considerably longer interval than
in the original, globally NFW reconstruction.

The first detected crossing time is $\tau_{\rm sc}\simeq1.05\times10^4$.
The 500- and 1000-shell estimates in Table~\ref{tab:numerical_diagnostics}
differ by about $0.05\%$, indicating modest resolution dependence on the
sampled domain rather than a continuum guarantee.
The oscillating configuration therefore provides an
example of a long-lived, but not indefinitely
shell-crossing-free, DBG solution.

The main numerical consistency checks are collected in
Table~\ref{tab:numerical_diagnostics}. The radial-equation residuals
remain small on the displayed slices, while the Jacobian diagnostics
quantify the finite interval of single-stream evolution.

\begin{table}[!tbp]
\centering
\caption{Numerical checks for the two examples on the sampled mass grids.
The Jacobian minimum and radial equation residual are evaluated over the
six displayed proper-time snapshots. Here
$D=(\partial R/\partial M)_t$ and
$W(R,M)$ is the analytical right-hand side of the radial equation.
Crossing times are grid-dependent estimates.}
\label{tab:numerical_diagnostics}
\begin{tabular}{@{}lcc@{}}
\toprule
Diagnostic & Collapse & Oscillations \\
\midrule
Mass domain & $[0.01,1]$ & $[0.002,0.998]$ \\
Number of shells & $500$ & $500$ \\
$\min_{\tau_j,M}D$ & $3.75$ & $50.35$ \\
$\max_{\tau_j,M}|\dot R^2-W|$ & $8.7\times10^{-18}$ & $5.1\times10^{-15}$ \\
$\min_M(E_u-E)$ & --- & $1.17$ \\
$\tau_{\rm sc}$ ($500$ shells) & --- & $10462.46$ \\
$\tau_{\rm sc}$ ($1000$ shells) & --- & $10457.20$ \\
\bottomrule
\end{tabular}
\end{table}

\FloatBarrier
\paragraph{Interpretation.}

The two numerical examples demonstrate complementary applications
of the reconstruction procedure.
The collapse example provides an outer branch solution approaching
a prescribed unstable EC endpoint asymptotically. The oscillating
example demonstrates how a stable Einasto equilibrium and a
finite time NFW-matched density configuration can be incorporated
into a single dynamical DBG model when an additional velocity
profile is specified.

The reconstruction does not imply that the Einasto or NFW profiles
are dynamical attractors. Instead, it provides a relativistic
framework in which prescribed density and velocity distributions
can be used to determine the shell dynamics and the corresponding
spacetime geometry.
The numerical results also emphasize the importance of checking
the physical radial Jacobian. A reconstruction may remain
well defined at the level of individual shell trajectories
even after the corresponding matter configuration ceases to
describe a regular, single-stream spacetime.
\FloatBarrier
\section{Summary and Conclusions}
\label{sec:conclusions}

We have formulated an inverse reconstruction scheme for the
Datta--Bondi--Gair family of spherically symmetric spacetimes with
vanishing radial pressure and tangential stress generated by conserved
angular momentum. Motivated by the Krasi\'nski--Hellaby method for LTB
dust, the construction uses density, velocity, equilibrium and
asymptotic endpoint data to determine the conserved shell functions and
shell-dependent phases. The novelty lies in this inverse formulation and
its metric regularity checks, rather than in a new matter model.

Two complementary configurations have been treated. First, imposing a
double root of the radial potential at a prescribed Einstein cluster
radius fixes $E(M)$ and $L(M)$. On the bound outer branch,
$4M<R_{\rm EC}<6M$, shells expand to a finite turnaround and approach
the unstable EC radius only in the infinite proper time limit. Second,
a stable static density profile with $R_0>6M$ fixes the equilibrium
angular momentum distribution, a finite-time density-velocity pair
then selects the energies and phases of finite amplitude oscillations.
The latter dynamics are reconstructed from the exact shell equation,
without assuming harmonic motion.

An essential component is the distinction between comoving coordinate
time and constituent-particle proper time. The radial reconstruction
$R(\tau,M)$ must be supplemented by the coordinate-transformation
function $K=(\partial\tau/\partial M)_t$. The physical Jacobian is
\[
D=(\partial R/\partial M)_t=(\partial R/\partial M)_\tau+\dot R K,
\]
while the energy density is $\rho=(4\pi R^2D)^{-1}$ and the
single-stream condition is $D>0$. This provides a consistent metric-level
interpretation of the numerical trajectories and identifies their
regular domain of validity.

We have illustrated the two constructions with a truncated Jaffe-type
asymptotic endpoint and with an Einasto equilibrium combined with an
NFW-matched finite-time profile. The first example excludes its genuine
central singularity. The second undergoes many radial oscillations
before shell crossing is detected near $\tau\simeq1.05\times10^4$ on
the sampled mass grid. These examples demonstrate admissible
solutions on specified domains; neither the occurrence of a generic
attractor nor global regularity follows from the reconstruction.

Several extensions would be useful. The prescribed angular-momentum
matter model offers tractable benchmark configurations for comparisons
with Einstein--Vlasov evolution, although a general kinetic
phase-space distribution is not reconstructed here
\cite{Andreasson2011}. Because the complete time-dependent metric is
available, horizons formation, photon propagation, redshift and time-delay calculations
could probe dynamical corrections to weak-field descriptions of
relativistic bound structures. Such applications would require careful
matching and appropriate low-compactness parameters
\cite{CarreraGiulini,Wiltshire2008}. A further motivation is the
study of boundaries between effectively bound regions and the
cosmological environment. In particular, the ``finite-infinity''
construction used in timescape cosmology
\cite{Wiltshire2007} poses a related but distinct problem: our
Schwarzschild matching at a comoving matter boundary does not itself
determine a finite-infinity surface or solve the cosmological
averaging problem. Extension to a cosmological exterior would be
needed to investigate that connection.

The inverse method therefore provides a controlled link between
relativistic tangential-pressure dynamics and physically prescribed
configurations, with explicit restrictions on stability, branch
selection and shell crossing.

\section*{Acknowledgements}
During the preparation of this work the author used the large language
models Claude Opus 5.5 (Anthropic) and ChatGPT-6 (OpenAI) to assist in
checking analytical derivations and in writing and debugging the numerical
code used for the examples and figures. All derivations and numerical
results were independently verified by the author, who takes full
responsibility for the content of this article.
\section*{Data availability}

The Python scripts and Jupyter notebook used to perform
the numerical reconstructions and generate the figures
presented in this work are publicly available at
\url{https://github.com/JanJOstrowski/DBG-Einstein-cluster-reconstruction}.

The repository includes the numerical implementation of
the proper time coordinate transformation, the physical
density and shell crossing diagnostics, and the input
parameters required to reproduce both examples. No observational
data sets were analysed.

\appendix
\section[Elliptic form of the proper-time quadrature]{Explicit elliptic form of the proper-time quadrature for $E<0$}
\label{Appa}
For the bound case ($E<0$, $\Lambda=0$) the proper-time separation between
two areal radii $R_A<R_B$ on a given shell is
\begin{equation}
\Delta\tau=\int_{R_A}^{R_B}\frac{\mathrm{d}\tilde R}
{\sqrt{W(\tilde R,M)}}\ ,\qquad
W=2E+\frac{2M}{R}-\frac{L^{2}}{R^{2}}+\frac{2ML^{2}}{R^{3}} \ .
\label{eq:A-dtau}
\end{equation}
Introducing $u=1/R$, so that $\mathrm{d}R=-\mathrm{d}u/u^{2}$,
\begin{equation}
W=P(u)\equiv 2E+2Mu-L^{2}u^{2}+2ML^{2}u^{3}
 =2ML^{2}\,(u-e_{1})(u-e_{2})(u-e_{3}) \ ,
\label{eq:A-cubic}
\end{equation}
and
\begin{equation}
\Delta\tau=\frac{1}{\sqrt{2ML^{2}}}
\int_{u_B}^{u_A}\frac{\mathrm{d}u}
{u^{2}\sqrt{(u-e_{1})(u-e_{2})(u-e_{3})}}\ ,\qquad u_{A,B}=1/R_{A,B}\ .
\label{eq:A-u}
\end{equation}

\paragraph{The physical interval.}
Order the roots as $e_{1}>e_{2}>e_{3}$. In terms of the three positive
roots $R_{i}<R_{\rm peri}<R_{\rm apo}$ of Sec.~\ref{sec:oscillation} one has
\begin{equation}
e_{1}=\frac{1}{R_{i}}\ ,\qquad e_{2}=\frac{1}{R_{\rm peri}}\ ,\qquad
e_{3}=\frac{1}{R_{\rm apo}}\ ,
\label{eq:A-roots}
\end{equation}
all strictly positive. Since the leading coefficient $2ML^{2}$ is
positive, $P(u)>0$ on $(e_{3},e_{2})$ and $P(u)<0$ on $(e_{2},e_{1})$.
The stable oscillation $R_{\rm peri}\le R\le R_{\rm apo}$ therefore
corresponds to
\begin{equation}
e_{3}\le u\le e_{2}\ ,
\label{eq:A-interval}
\end{equation}
and on this interval it is convenient to write the radical in manifestly
real form,
\begin{equation}
\sqrt{(u-e_{1})(u-e_{2})(u-e_{3})}
=\sqrt{(e_{1}-u)(e_{2}-u)(u-e_{3})}\ .
\end{equation}

\paragraph{Transformation to Legendre normal form.}
Set
\begin{equation}
u=e_{3}+(e_{2}-e_{3})\sin^{2}\theta\ ,\qquad
\mathrm{d}u=2(e_{2}-e_{3})\sin\theta\cos\theta\,\mathrm{d}\theta\ ,
\label{eq:A-sub}
\end{equation}
so that $\theta=0$ at $u=e_{3}$ ($R=R_{\rm apo}$) and $\theta=\pi/2$ at
$u=e_{2}$ ($R=R_{\rm peri}$). The factors of the cubic become
\begin{equation}
e_{2}-u=(e_{2}-e_{3})\cos^{2}\theta\ ,\quad
u-e_{3}=(e_{2}-e_{3})\sin^{2}\theta\ ,\quad
e_{1}-u=(e_{1}-e_{3})\bigl(1-m\sin^{2}\theta\bigr)\ ,
\end{equation}
with modulus
\begin{equation}
m=\frac{e_{2}-e_{3}}{e_{1}-e_{3}}\in(0,1)\ .
\label{eq:A-m}
\end{equation}
Hence
$\sqrt{(e_{1}-u)(e_{2}-u)(u-e_{3})}
=(e_{2}-e_{3})\sqrt{e_{1}-e_{3}}\,\sin\theta\cos\theta
 \sqrt{1-m\sin^{2}\theta}$,
and the $\sin\theta\cos\theta$ cancels against $\mathrm{d}u$. Writing
\begin{equation}
u(\theta)=e_{3}\bigl(1-n\sin^{2}\theta\bigr)\ ,\qquad
n=-\frac{e_{2}-e_{3}}{e_{3}}<0\ ,
\label{eq:A-n}
\end{equation}
Eq.~\eqref{eq:A-u} becomes
\begin{equation}
\Delta\tau=\frac{2}{e_{3}^{2}\sqrt{2ML^{2}(e_{1}-e_{3})}}
\int_{\theta_B}^{\theta_A}
\frac{\mathrm{d}\theta}
{\bigl(1-n\sin^{2}\theta\bigr)^{2}\sqrt{1-m\sin^{2}\theta}}\ ,
\label{eq:A-legendre}
\end{equation}
with limits fixed by
\begin{equation}
\sin^{2}\theta_{X}=\frac{u_{X}-e_{3}}{e_{2}-e_{3}}
=\frac{1/R_{X}-e_{3}}{e_{2}-e_{3}}\ ,\qquad X=A,B\ .
\label{eq:A-limits}
\end{equation}
Note that $n<0$ because $e_{2}>e_{3}>0$, so $1-n\sin^{2}\theta\ge 1$
throughout: the characteristic lies in the circular case and the
integral has no pole, requiring no principal-value prescription.

\paragraph{Reduction to Legendre elliptic integrals.}
Let
\begin{equation}
J_{2}(\phi)\equiv\int_{0}^{\phi}
\frac{\mathrm{d}\theta}
{\bigl(1-n\sin^{2}\theta\bigr)^{2}\sqrt{1-m\sin^{2}\theta}}\ .
\end{equation}
Differentiating
$G(\phi)=\sin\phi\cos\phi\sqrt{1-m\sin^{2}\phi}\,/\,(1-n\sin^{2}\phi)$
and decomposing the result in powers of $(1-n\sin^{2}\phi)$ gives the
exact reduction
\begin{equation}
J_{2}(\phi)=\frac{n^{2}G(\phi)
-\bigl(2mn+2n-n^{2}-3m\bigr)\,\Pi(n;\phi\,|\,m)
-(m-n)\,F(\phi\,|\,m)-n\,E(\phi\,|\,m)}
{2\,(1-n)\,(m-n)}\;
\label{eq:A-J2}
\end{equation}
in terms of the incomplete elliptic integrals of the first, second and
third kinds. 
The proper time is then
\begin{equation}
\Delta\tau(M)=\frac{2}{e_{3}^{2}\sqrt{2ML^{2}(e_{1}-e_{3})}}
\bigl[J_{2}(\theta_{A})-J_{2}(\theta_{B})\bigr]\ .
\label{eq:A-final}
\end{equation}
For the half-period one has $\theta_{A}=\pi/2$, $\theta_{B}=0$, and
since $G(\pi/2)=G(0)=0$,
\begin{equation}
H(M)=\frac{-\bigl(2mn+2n-n^{2}-3m\bigr)\Pi(n\,|\,m)
-(m-n)K(m)-nE(m)}
{\,e_{3}^{2}\,(1-n)(m-n)\sqrt{2ML^{2}(e_{1}-e_{3})}\,}\ ,
\qquad T(M)=2H(M)\ ,
\label{eq:A-period}
\end{equation}
with $K$, $E$, $\Pi$ the complete integrals.

\paragraph{Degenerate limit.}
The asymptotic collapse of Sec.~\ref{sec:ec_final_state} is the limit in which
the two inner roots merge, $e_{1}\to e_{2}=1/R_{\rm EC}$, with
$e_{3}=1/R_{\rm ta}$. Equation \eqref{eq:A-m} then gives $m\to1$, where
$K(m)$ and $\Pi(n|m)$ diverge logarithmically. Equation
\eqref{eq:A-period} therefore reproduces analytically the infinite
proper time required to reach a double root, i.e. the logarithmic behaviour derived near the double root in Sec.~\ref{sec:ec_final_state}, and the two
constructions of Sec.~4 are the $m<1$ and $m\to1$ cases of a single
quadrature.

\paragraph{Remark on numerical evaluation.}
In practice it is preferable to evaluate \eqref{eq:A-u} in Carlson's
symmetric form, which is free of the branch and ordering conventions of
the Legendre reduction and which degrades gracefully as $m\to1$ or as
any two roots approach each other; implementations are available in
standard libraries. The Legendre form \eqref{eq:A-J2}--\eqref{eq:A-period}
is given here because it exhibits explicitly which elliptic integrals
control the shell period, and because it makes the $m\to1$ degeneration
transparent.

\bibliography{references}

\end{document}